\documentclass[aps,pra,twocolumn]{revtex4-2}
\usepackage{amsmath}
\usepackage{color}
\usepackage{graphicx}
\usepackage{float}
\usepackage{braket}
\usepackage{xcolor}

\begin{document}

\title{Ladder of Loschmidt anomalies in the deep strong-coupling regime of a\\  qubit-oscillator system}
\author{J.~M. Betancourt}
\affiliation{Departamento de F{\'i}sica, Universidad de los Andes, A.A. 4976, Bogot{\'a} Distrito Capital, Colombia}
\author{F.~J. Rodr\'iguez}
\affiliation{Departamento de F{\'i}sica, Universidad de los Andes, A.A. 4976, Bogot{\'a} Distrito Capital, Colombia}
\author{N.~F. Johnson}
\affiliation{Department of Physics, George Washington University, Washington D.C. 20052, U.S.A.}
\author{L. Quiroga}
\affiliation{Departamento de F{\'i}sica, Universidad de los Andes, A.A. 4976, Bogot{\'a} Distrito Capital, Colombia}
\email{lquiroga@uniandes.edu.co}

\date{\today}

\begin{abstract}
We uncover a remarkably regular array of singularity-like structures within the deep strong-coupling limit of qubit-oscillator (e.g. light-matter) systems  described by the quantum Rabi model, as a function of time and coupling strength. These non-analytic anomalies in the Loschmidt amplitude (echoes) suggest the existence of new forms of dynamical phase transition within this deep strong-coupling regime. The key feature whereby the initial state collapses into orthogonal states at select values of the interaction strength and select times, may be used to enhance -- or attack -- quantum information processing or computation schemes that rely on removing -- or retaining --  a given quantum state.

\end{abstract}
\maketitle

\section{Introduction}
Qubit-oscillator systems (e.g. confined light-matter interactions or spin-nanomechanical resonators) play a key role in physics and its applications, including the burgeoning field of quantum computing~\cite{nielsen_chuang_2010} as well as quantum optics~\cite{cohen1989photons}, molecular physics~\cite{Thanopulos2004} and hybrid quantum devices~\cite{Li2015}. This motivates the search for a deeper theoretical understanding of their quantum dynamics. The quantum Rabi model (QRM)~\cite{Rabi1936} describes the fundamental building block for such systems -- a single two-level qubit (e.g. atom, quantum dot or NV center) coupled to an oscillator (e.g. photon or phonon) mode -- whereas the Dicke model~\cite{Dicke1954} generalizes this to multiple qubits. Although there is an exact solution for the
spectrum of the QRM~\cite{Braak2011, Wolf2012}, the resulting expression is difficult to manipulate for practical dynamical applications.

Perturbative methods have been developed for different values of the qubit-oscillator (e.g. light-matter) coupling strength $g$, leading to a classification of the system into regimes depending on the value of $g$~\cite{Rossatto2017}.  When $g$ is considerably smaller than the natural frequencies of the qubit and the bosonic mode, i.e. the weak coupling regime, the QRM is well described by the exactly solvable Jaynes-Cummings model~\cite{Jaynes1963}. When $g$ is a significant fraction of these frequencies, the system lies in the strong coupling (SC) or ultrastrong coupling (USC) regimes. When the coupling strength becomes larger than the natural frequencies of the system, the system lies in the deep strong coupling (DSC) regime. Although experiments have been performed in the SC regime for decades via cavity QED ~\cite{auffeves2014strong}, recent  advances in achieving the USC ~\cite{Todorov2010, Forn-Diaz2010, Niemczyk2010} and DSC~\cite{Yoshihara2017, Langford2017, Bayer2017} experimentally have brought these regimes to the forefront of attention.

The current capability to experimentally vary the strength of the coupling $g$ over a wide range of values raises the possibility of exploring non-analytic behaviors that point toward new phase transitions~\cite{Gould2010}. Such behaviors could also arise as a function of time in a driven system, i.e. a dynamical quantum phase transition (DQPT)~\cite{Heyl2019}. One of the key quantities in the study of DQPTs is the Loschmidt amplitude (LA) which is the overlap between the initial quantum state of the system and the time-evolved one. Its zeros play the same role in the complex time plane as the Yang-Lee zeros of the partition function do in the complex temperature plane~\cite{Heyl2013}. Sharp, non-analytic features in its time-evolution suggest dynamical phase transitions~\cite{Heyl2019}. Recent theoretical work on the single qubit QRM has shown several quantum phase transition behaviors in the off-resonant limiting case where the ratio between the qubit splitting energy and the mode goes to infinity at the same time as $g$ goes to infinity~\cite{Hwang2015}. In this limit,  the ground-state of the QRM undergoes a second-order phase transition whereas excited states show signatures of an excited state quantum phase transition~\cite{Puebla2016}. Experimental realizations with a single trapped ion have already been reported~\cite{Cai2021}, and a DQPT has been proposed in the off-resonant limit~\cite{Puebla2020}. More generally, studies across different types of qubit-oscillator systems have suggested that the ultra-strong coupling regime is relevant to wide a range of fundamental physics phenomena such as virtual photon accessibility in the Casimir effect~\cite{Kockum2019, Stange2021} and novel  applications including optomechanical nanosystems~\cite{Stannigel2012, Kotler2021, Schmidt2010}.

This paper shows that deep strong coupling generates a new regime of non-analytic features  suggestive of dynamical phase transitions, even for the simplest of systems (QRM). Our full numerical calculations show that approximate zeros of the LA (echoes) emerge in a staircase structure as a function of  the coupling strength $g$ and time $t$ after the onset of this coupling. In addition to being relevant to the fundamental quantum speed limit for information processing, i.e. minimal time for the evolution to an orthogonal state~\cite{Vaidman1992, Levitin2009}, these orthogonality-points in $(g,t)$ space can provide well-defined stages in a quantum computation or information processing scheme. For situations where they are undesirable, our findings provide the combinations of $g$ and $t$ to avoid.

This paper is organized as follows. We begin by introducing the system Hamiltonian in Sec.~\ref{sec: model}, as well as the Loschmidt echo for an initial state with no photons and no qubit excitation. In Sec.~\ref{sec: singularities}, we show numerical results where Loschmidt echo singularities appear. In addition, an approximation using the zero qubit splitting case and an equation of motion involving the derivatives of the echo are derived.  Concluding remarks are given in Sec. \ref{sec: conclusion}.

\section{Hamiltonian model} \label{sec: model}
We consider arguably the simplest example of a driven
qubit-oscillator system: a two-level system with level splitting $\omega_0$ described by Pauli operators $\hat{\sigma}_z$, $\hat{\sigma}^{+}$ and $\hat{\sigma}^{-}$, coupled with a time dependent strength $g(t)$ to a single boson mode (QRM) with frequency $\omega_c$:
\begin{equation}\label{e1}
\hat{H}(t)=\omega_c\hat{a}^{\dag}\hat{a}+\frac{\omega_0}{2}\hat{\sigma}_z+g(t)(\hat{a}^{\dag}+\hat{a})(\hat{\sigma}^{+}+\hat{\sigma}^{-}),
\end{equation}
where $\hbar=1$. The operators $\hat{a}^{\dag}$ $\hat{a}$ denote boson ladder operators, which for simplicity we call photon operators.

There is a controversy in the literature about the contribution of the quadratic term of vector potential $A$, related with the quantum phase transitions for the Dicke model \cite{Bernardis2018}. By contrast, our paper is focused on the dynamical overlap between the initial state and the state later in time -- it is not focused on this transition. Therefore, we eliminate a possible $A^2$ term by performing a Bogoliugov-Hopfield transformation on the photon operators.

This Hamiltonian commutes with the parity operator $\hat{P}=e^{i\pi (\hat{a}^{\dag}\hat{a}+\frac{1}{2}(1+\hat{\sigma}_z))}$. Given an initial state with zero excitations in both qubit and oscillator sub-systems $|\psi(0)\rangle=|0,0\rangle$ (i.e. even parity) the time-evolution remains confined to the even Hilbert subspace. Hence a basis can be chosen as $\{ | n,m\rangle \}$ with $n+m=$ even and $n=0,1,2,...$ denoting the number of boson excitations whereas $m=0,1$ corresponds to the number of qubit excitations. We can denote the basis vectors simply by $\{ | n\rangle \}\equiv\{ | n,m\rangle \}$, noting that $m=0$ when $n$ is even whereas $m=1$ otherwise. Hence Eq.~\eqref{e1} becomes:
\begin{equation}\label{e4}
\hat{H}(t)=\omega_c\hat{a}^{\dag}\hat{a}-\frac{\omega_0}{2}e^{i\pi \hat{a}^{\dag}\hat{a}}+g(t)(\hat{a}^{\dag}+\hat{a}).
\end{equation}
This form explicitly shows the strong anharmonicity, mediated by the qubit, of the oscillator sub-system.
The LA
\begin{eqnarray}\label{e20}
{\cal L}(t)=\langle \psi(0)|\hat{U}(t,0)|\psi(0)\rangle,
\end{eqnarray}
calculates the fidelity of the time-evolved state $|\psi(t)\rangle=\hat{U}(t,0)|\psi(0)\rangle$ with respect to the initial state $|\psi(0)\rangle$. We calculate the final state $|\psi(t)\rangle$ under Eq.~\eqref{e1} given the initial ground state $|\psi(0)\rangle\equiv|0,0\rangle$. Furthermore, throughout the paper we consider the simplest quenching case in which the qubit-cavity coupling strength jumps suddenly from 0 to a constant value $g > 0$ at time 0, and then maintains that constant value for time $t$.
Although the Rabi model is integrable~\cite{Braak2011}, no analytical results are available for the case of interest here, i.e. the non-equilibrium resonant Rabi model.


\begin{figure}
	\centering
	\includegraphics[width = \linewidth]{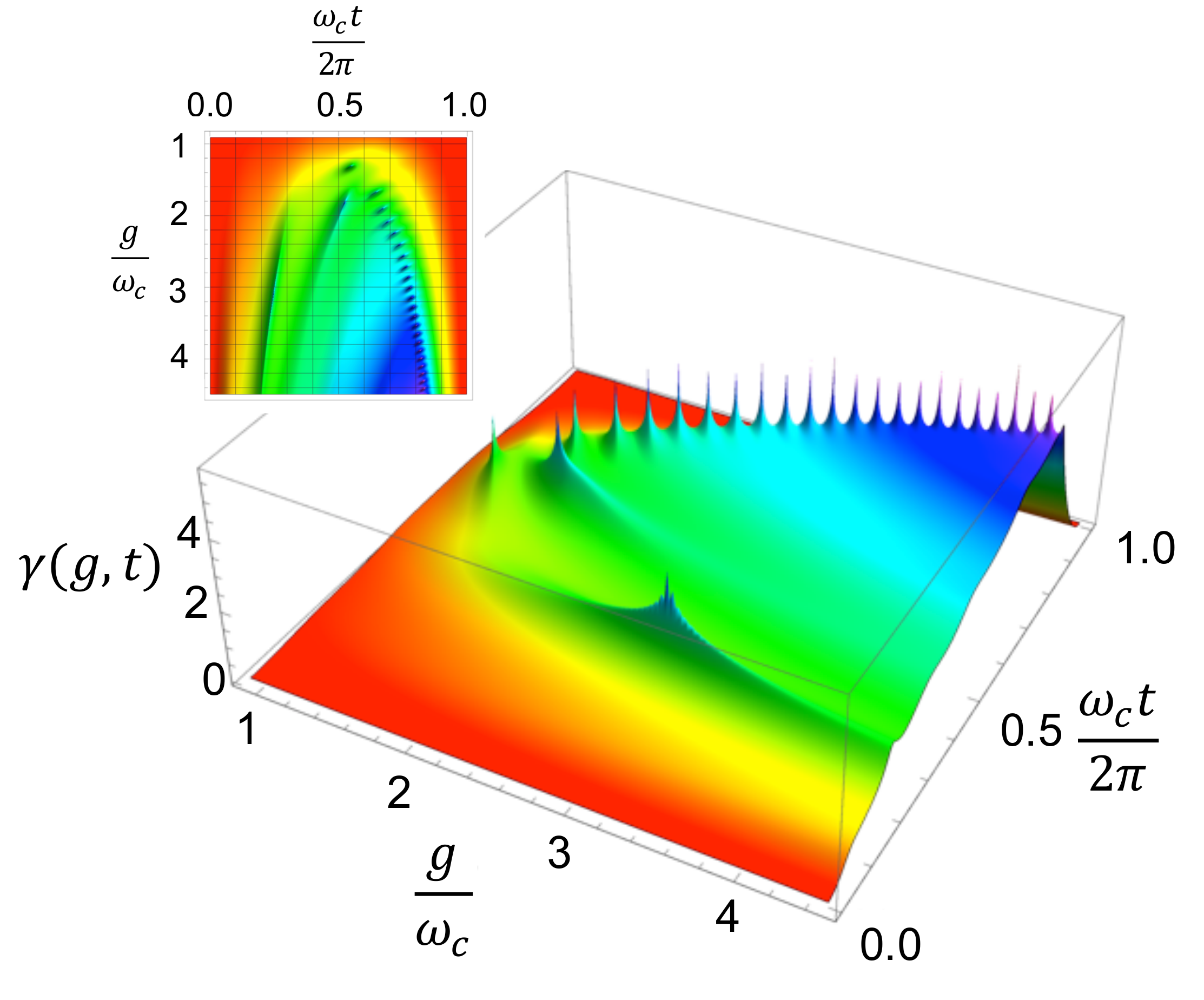}	
	\caption{A ladder of singularity peaks in the Loschmidt echo emerges in the deep strong coupling (DSC) regime, as a function of the coupling $g$ and time $t$ after the coupling onset. Results are shown in two dimensions (2D) and three dimensions for the resonant case, $\omega_c=\omega_0$, and a single cycle time evolution, $\frac{\omega_c t}{2\pi} \leq 1$.}
	\label{fig:rates}
\end{figure}

\begin{figure}
	\centering
	\includegraphics[width = \linewidth]{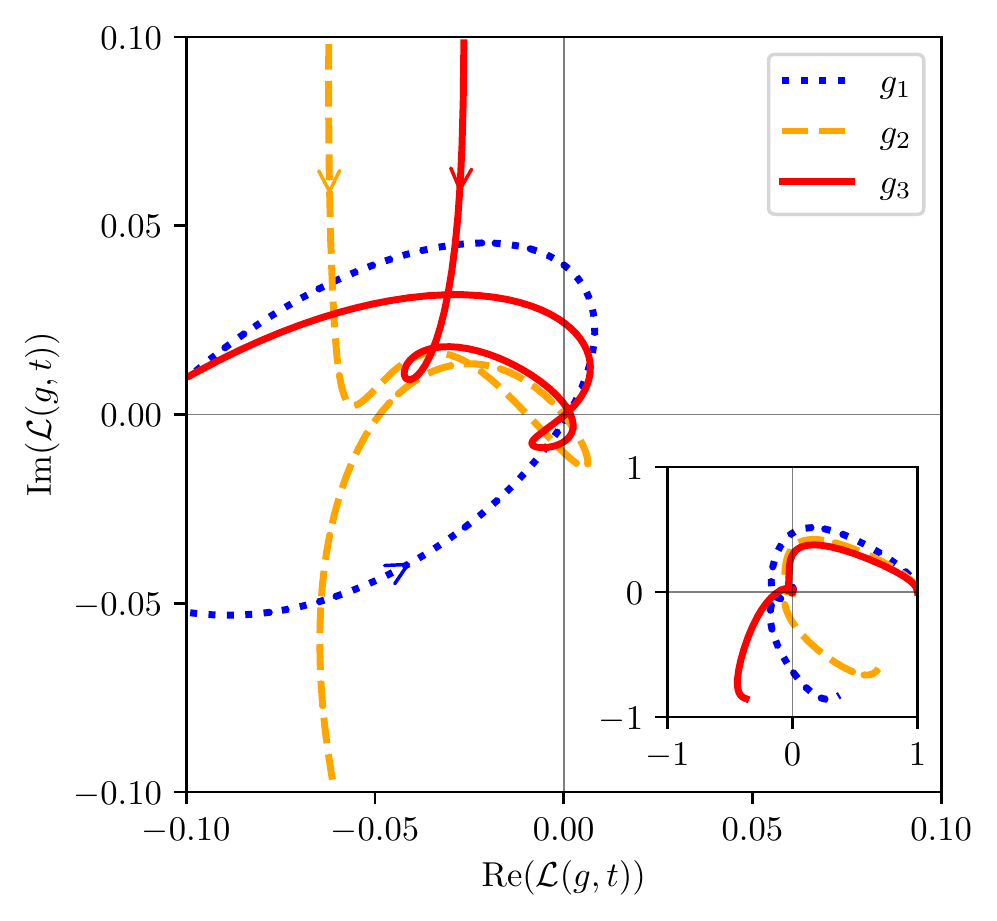}
	\caption{Parametric plots of $\mathcal{L}(g,t)$ for the coupling values $g$ where the first three zeros of $\mathcal{L}$ arise (the $n$-th lowest value is denoted by $g_n$). Arrows indicate the LA evolution as time progresses. Changes of LA chirality are evident for $g_2$ and $g_3$. The inset: LA full evolution for $0 \leq \frac{\omega_c t}{2\pi} \leq 1$ in a full scale view, for the same $g_n$ values.}
	\label{fig:complex_echo}
\end{figure}

We benchmark tested our full numerical calculations using the $\omega_0\rightarrow 0$ limit for which the exact analytic solution is known ~\cite{Gilbey1966, Galindo1991}, noting that our calculations can be generalized to other time-dependent quenchings \cite{Gomez-Ruiz2018}. Although this simple limiting case does not show any signature of a dynamical phase transition, the situation changes drastically with a finite qubit splitting $\omega_0=\omega_c$.
For convenience, we use a logarithmic scale by defining a Loschmidt echo rate (LER) exponent as~\cite{Heyl2019} $\gamma(g,t)=-\log_{10}|{\cal L}(g,t)|$
whereas corresponding exponents can be defined for derivatives with respect to $t$ and $g$: $\beta(g,t)=-\log_{10}\left |\frac{\partial{\cal L}(g,t)}{\partial t}\right |$ and $\chi(g,t)=-\log_{10}\left |\frac{\partial{\cal L}(g,t)}{\partial g}\right |$.

\begin{figure}
	\centering
	\includegraphics[width = \linewidth]{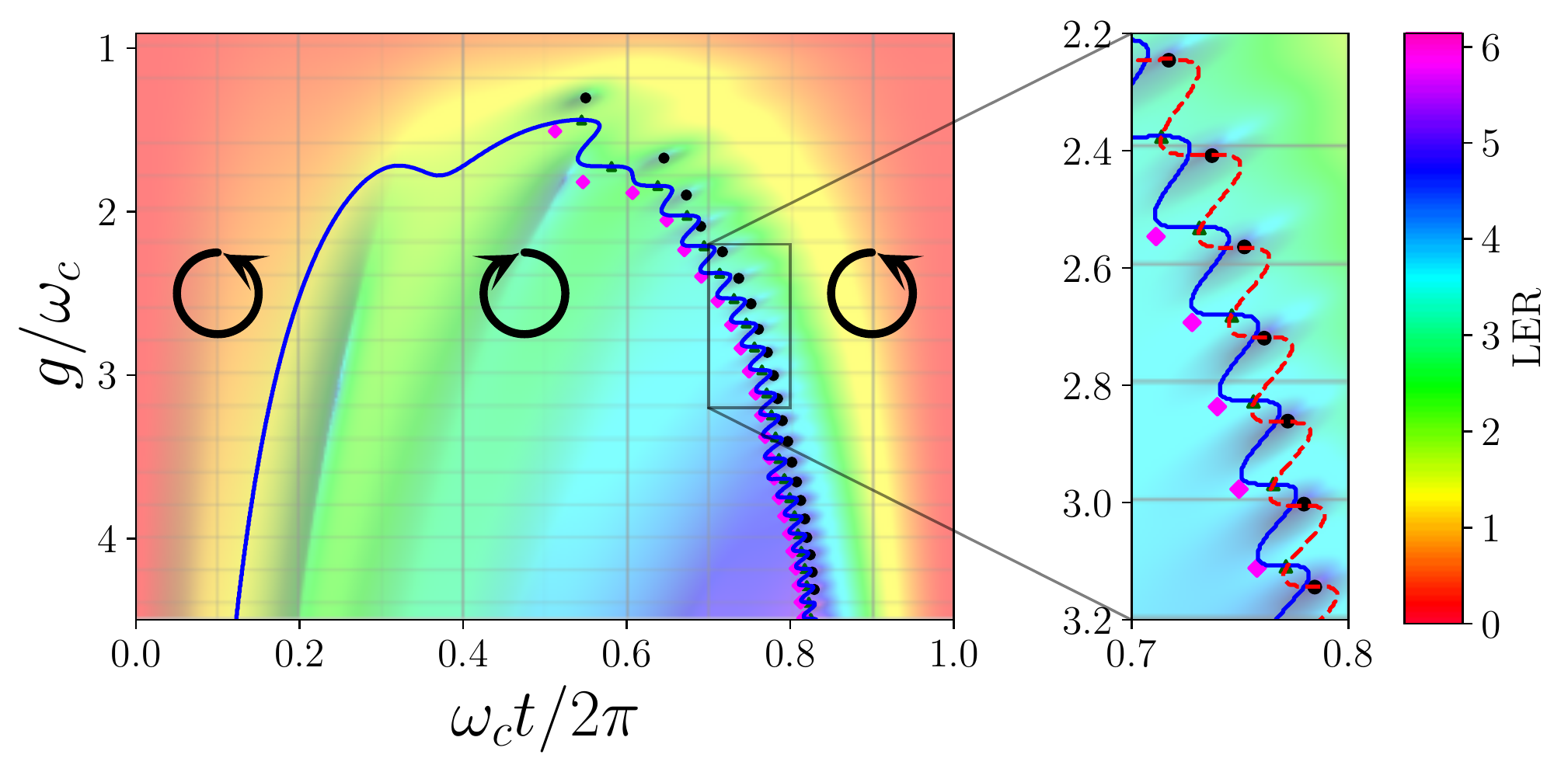}
	\caption{Boundary between regions of clockwise and counter- clockwise rotation of the time derivative of the LA (blue solid line) in the complex plane shown over a 2D projection of Fig 1. The boundary corresponds to $\partial_t \arg(\partial_t \mathcal{L}(g,t)) = 0$. Singularities of $\gamma$ (black dots), $\beta$ (green triangles) and $\chi$ (magenta diamonds) are also shown. Changes in chirality are more clearly seen in the right panel, where a red dashed line indicating $\partial_t \arg(\mathcal{L}(g, t)) = 0$ is included.}
	\label{fig:rotation}
\end{figure}



\section{Loschmidt singularities} \label{sec: singularities}
In this section we study the time-dependent Loschmidt echo by solving the quantum propagator numerically for the Rabi model. For a pulsed coupling of constant height $g$, the dynamics are contained within the complete set of even parity eigenstates $|E_m\rangle$ with energy $E_m$ for the QRM Hamiltonian (Eq.~\eqref{e1}) where ${\cal L}(g,t)=\sum_{m}e^{-iE_mt}|\langle E_m|\psi(0)\rangle|^2$. Fig. \ref{fig:rates} shows our results for LER ($\gamma(g,t)$) at $\omega_0=\omega_c$ and $|\psi(0)\rangle=|0,0\rangle$, across a wide region of the DSC $(g,t)$ plane. This regime -- which is largely unexplored -- can only be understood in those systems, if non-perturbative methods are correctly included in the calculation. We find that using a basis set of up to 750 photon states guarantees convergence. Although a smooth continuum occupies the left half of the one-period evolution, a ladder of localized singularities arises for $\gamma(g,t)$ in the right half (see Appendix \ref{app: chirality} for similar features in $\beta(g,t)$ and $\chi(g,t)$). We emphasize that this ladder of LER singularities has nothing to do with quenching through an equilibrium quantum phase transition, since the QRM at resonance doesn't show the emergence of any kind of equilibrium phase transition. Therefore, we conclude that this non-analytic dynamical behavior of the LER at select points in the $(g,t)$ plane, can be understood as a DQPT when quenching from the uncoupled qubit-oscillator ground state to a final DSC coupled system -- regardless of the finite size of the matter sub-system.

So far we have pointed out the existence of a ladder of LA echoes. Deeper insight is gained by inspecting the trajectories of the full LA in the complex plane. In Fig. \ref{fig:complex_echo}, we show these trajectories for the first three values of $g$ where zeros can be located, namely, $g_1 = 1.30$ (blue dotted line), $g_2 = 1.67$ (orange dashed line) and $g_3 = 1.90$ (red solid line). It is important to note that the chosen values of $g$ have been achieved experimentally~\cite{Yoshihara2017, Langford2017, Bayer2017}. These trajectories seem to present increasingly complex winding behaviors, suggesting that information about the structure of the LA might be embedded into these geometric aspects. This leads us to conclude that the ladder of LA zeros is intimately connected with DQPTs, which are well known to have no local order parameter. Instead, DQPTs have been associated with dynamical {\em topological} order parameters which can be extracted from the Pancharatnam phase of the LA itself \cite{Ding2020}. For critical translational invariant systems, the dynamical topological order parameter is quantized as an integer number that jumps at critical times. We do not know of any previous reports of such signatures in a system such as the QRM that does {\em not} have an associated length scale.

The LA exhibits highly nontrivial behavior in terms of its rotation in the complex plane, which can be captured by analyzing the temporal evolution of $\arg (\partial_t \mathcal{L}(g,t))$. Specifically, $\partial_t \arg (\partial_t \mathcal{L}(g,t)) > 0$ and $\partial_t \arg (\partial_t \mathcal{L}(g,t)) < 0$ correspond to anti-clockwise and clockwise rotation of the LA velocity, respectively. Fig. \ref{fig:rotation} shows the rotation directions of the LA and its time derivative in different regimes, and the location of the singularities of $\gamma$, $\beta$ and $\chi$. Notice that the singularities of $\gamma$ must lie on the red (dashed) boundary, whereas the singularities of $\beta$ must always lie on both boundaries (see Appendix \ref{app: chirality} for details).
In the DSC regime, we find alternating regions of clockwise and anti-clockwise rotations.
Fig. \ref{fig:rotation} reveals a rich dynamical phase diagram with additional topological phase transitions which are a different class from those previously reported for the off-resonant QRM. The phases separated by these topological transitions are distinguished from each other by this feature, and each can in principle be assigned a topological number. Interestingly, the boundary between these regions seems to follow the location of the discrete ladder of LA zeros for higher values of $g$ (Fig. \ref{fig:rotation} right panel) indicating a hidden relationship between changes in this rotational behavior and the emergence of singularities. Moreover, zeros in the LA derivatives with respect to time ($\beta(g,t)$) and coupling ($\chi(g,t)$) fall on this same undulating border that separates different LA-chirality regions. This is a key signature of the emergence of a topological dynamical phase transition.

\begin{figure}
	\centering
	\includegraphics[width = \linewidth]{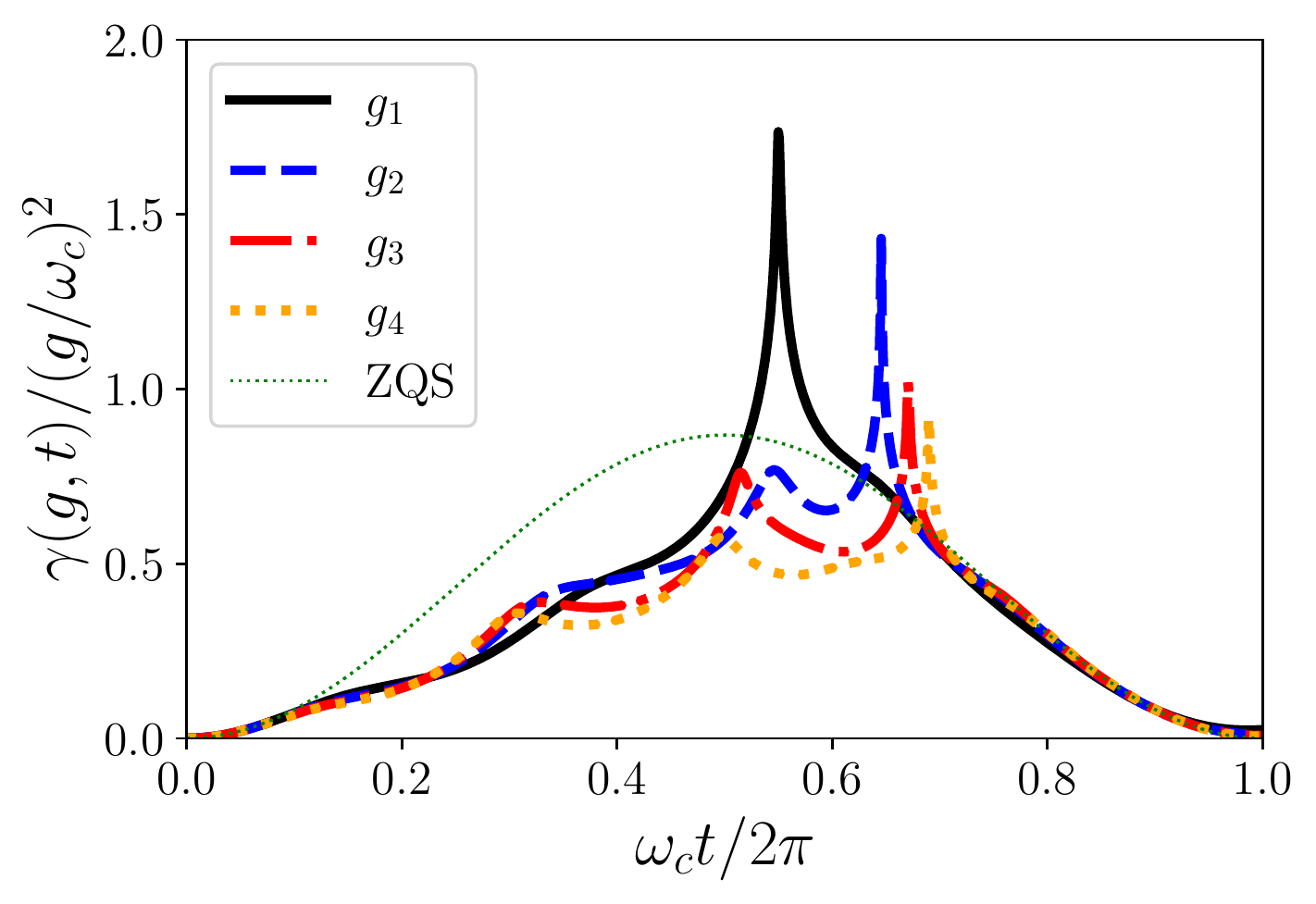}
	\caption{ Numerical and ZQS results for $\gamma(g,t)/(g/\omega_c)^2$, for coupling strength $g$ values corresponding to the first four LA zeros. In the trailing edge of the coupling pulse, there is a collapse of the exact numerical results onto the curve for the ZQS (Eq.~\eqref{e771}).}
	\label{fig:approx}
\end{figure}

For the LA and LER exponents, we find a near perfect collapse of numerical values of $\gamma(g,t)/g^2$ for the QRM, with the corresponding analytical result for the zero qubit splitting (ZQS) limit result given by
\begin{eqnarray}\label{e221}
{\cal L}_0(g,t)=e^{i\theta(g,t)}e^{-2\left (\frac{g}{\omega_c}\right )^2{\rm sin}^2\left ( \frac{\omega_c t}{2} \right )\ .} \end{eqnarray}
This yields
\begin{eqnarray}\label{e771}
\gamma_0(g,t)/g^2\sim {\rm sin}^2\left ( \frac{\omega_c t}{2} \right )
\end{eqnarray}
as a universal value which is independent of the specific coupling strength $g$ (see Appendix \ref{app: ZQS} for the expression of $\theta(g,t)$ and additional details). In the trailing edge of the coupling pulse, there is good agreement between the exact numerical results and the expression calculated from the Loschmidt echo rate for a ZQS system $|{\cal L}_0(g,t)|$ in Eq.~\eqref{e221}, especially after the kink in $\gamma(g,t)/g^2$ (see Fig. \ref{fig:approx}). This collapse suggests that the anharmonicity induced by the qubit in the effective oscillator in Eq.~\eqref{e4}, fades away for long pulse durations that place the system to the right of the wavy chirality border in Fig. \ref{fig:rotation}. This raises the possibility of designing tailored pulses for which the LA and the corresponding exponent become {\em identical} in the long time pulse limit for the highly anharmonic resonant qubit-boson case and for the simple harmonic ZQS system.

Further insight into the emergence of this LA ladder comes from the following exact relationship for an arbitrary quantum system  \cite{Campo2016}
\begin{eqnarray}\label{z66}
{\cal L}(t)={\cal L}(t-\tau){\cal L}(\tau)+M(t,\tau)
\end{eqnarray}
where the memory term is
\begin{eqnarray}\label{z6}
M(t,\tau)=\langle \psi_0|\hat{U}(t,\tau)\, \hat{Q}\, \hat{U}(\tau,0)|\psi_0\rangle
\end{eqnarray}
and $\hat{Q}=1-|\psi_0\rangle \langle \psi_0|$. Using Eqs.~\eqref{e20} and \eqref{z66}, we obtain an equation of motion for the LA
\begin{eqnarray}
g\frac{\partial\,{\cal L}}{\partial\,g}-t\frac{\partial\,{\cal L}}{\partial\,t}=-\frac{it}{2}{\cal L}(g,t)
+{\cal G}(g,t)
\label{Eq:z18}
\end{eqnarray}
where
\begin{align*}
{\cal G}(g,t)=&it\sum_{n=1}^{\infty}\left [ \frac{(-1)^{n+1}-1}{2}+n\right ] \\
  &\int_0^1 dx\, \langle \psi_0|e^{-ix\hat{H}(g)t}|n\rangle \langle n|e^{-i(1-x) \hat{H}(g)t}|\psi_0\rangle
\end{align*}
and $\{ | n\rangle \}\equiv\{ | n,m\rangle \}$ (see Appendix \ref{app: EOM} for derivation). The staircase nature of the anomalies (i.e. LA zeros) as a function of $g$ and $t$, reflects the fact that the LA equation of motion resembles a generalized Burger's shock-wave equation in which $\frac{\partial {\cal L}}{\partial t}-(g/t)\frac{\partial {\cal L}}{\partial g}$ contains a convolution of ${{\cal L}(g,t)}$ over time plus correction terms. This suggests the propagation of short-lived, modified shock waves in ${\cal L}(g,t)$ as a function of $g$ and $t$, as observed.

\section{Concluding remarks} \label{sec: conclusion}
Our findings have further implications. First, this discrete staircase of LA-zeros not only appears for the QRM, but it also shows signs of existing for multi-qubit systems (Dicke model) and with more complicated time varying couplings~\cite{Gomez-Ruiz2016}. Although a comprehensive treatment of the scaling of such complex systems remains to be explored, our results open up a path to tailoring zeros of the LA using a time-dependent interaction and by increasing the number of qubits.

Second, photon states lacking the vacuum component are predicted to present highly non-classical features, such as negative values of the Wigner function and Schr\"{o}dinger-cat state generation~\cite{Ashhab2010}. Our findings demonstrate that a simple QRM (e.g. single qubit-photon) system is an ideal platform for realizing such non-classical light states, simply by varying the coupling $g$ and time since onset $t$.

Third, the key feature whereby the initial state collapses into orthogonal states in a punctuated way as a function of $g$ and $t$, may be used to enhance -- or attack -- future quantum information processing or computation implementations that rely on removing -- or retaining --  a given quantum state. This finding could motivate new schemes that exploit this fact for erasing memory of an initial state. However, it also flags a potential vulnerability to external attack in quantum information processing schemes that require maintaining some memory (i.e. non-zero vacuum component) of the initial state.

Finally, we note that  decoherence should have a negligible effect on the ladder of LA anomalies reported here, since they occur within a single cavity cycle scale of time and in the DSC regime, placing the system in an ultrafast regime where the unitary treatment discussed above should be appropriate.

\begin{acknowledgments}
We thank F.J. G\'omez-Ruiz for discussions. J.M.B, F.J.R. and L.Q. are thankful to Facultad de Ciencias-UniAndes project ``{\it Excited State Quantum Phase Transitions in Driven Models - Part II: Dynamical QPT}" (2019-2020) for partial financial support. N.F.J. was funded by the AFOSR through Grants No. FA9550-20-1-0382 and No. FA9550-20-1-0383.
\end{acknowledgments}

\appendix
\section{Chirality and singularities of $\beta$ and $\chi$} \label{app: chirality}
\begin{figure}
	\centering
	\includegraphics[width = \linewidth]{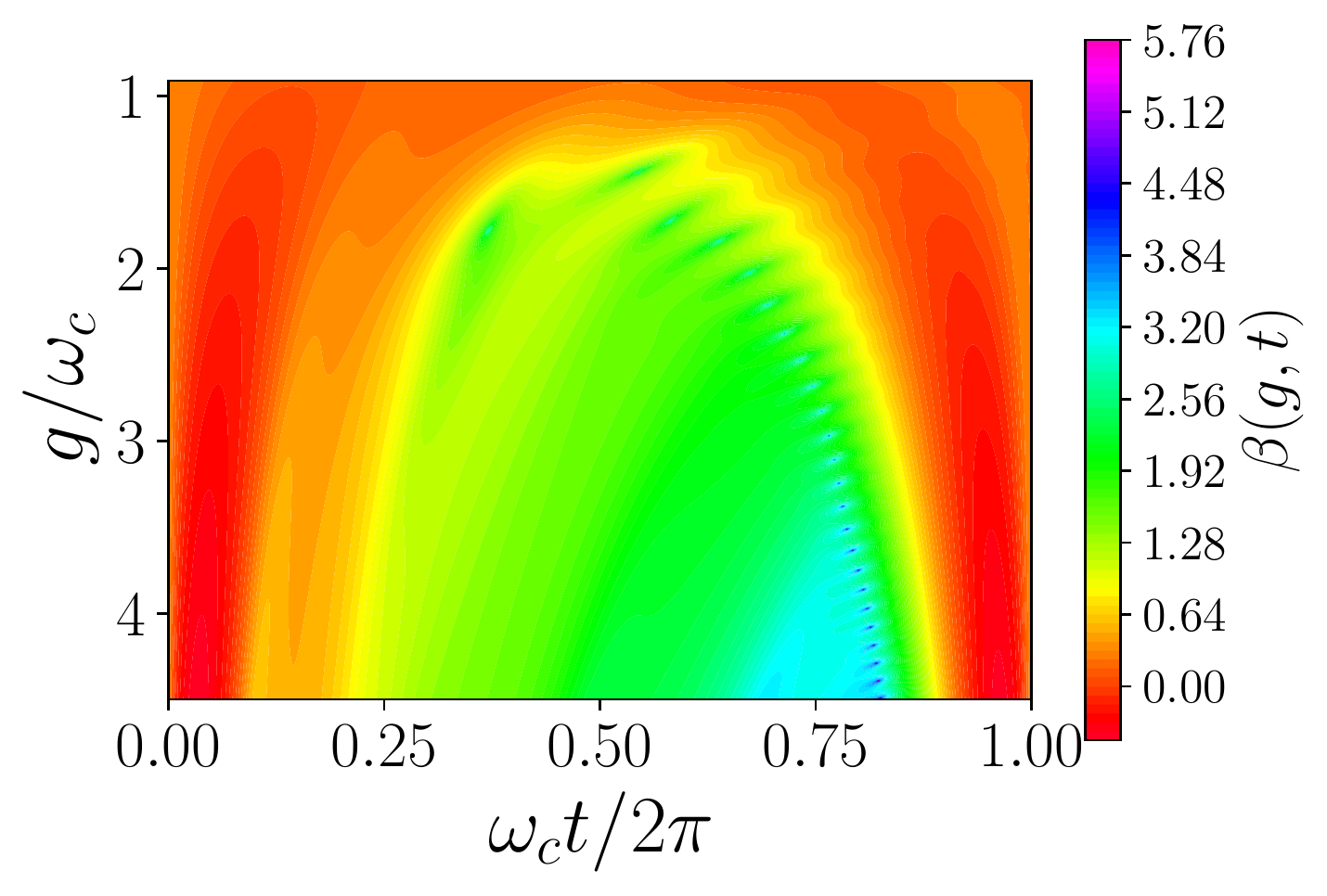}
	\caption{Values of $\beta(g,t)$ for the analyzed values of coupling strength and evolution time.}
	\label{fig:beta}
\end{figure}

\begin{figure}
	\centering
	\includegraphics[width = \linewidth]{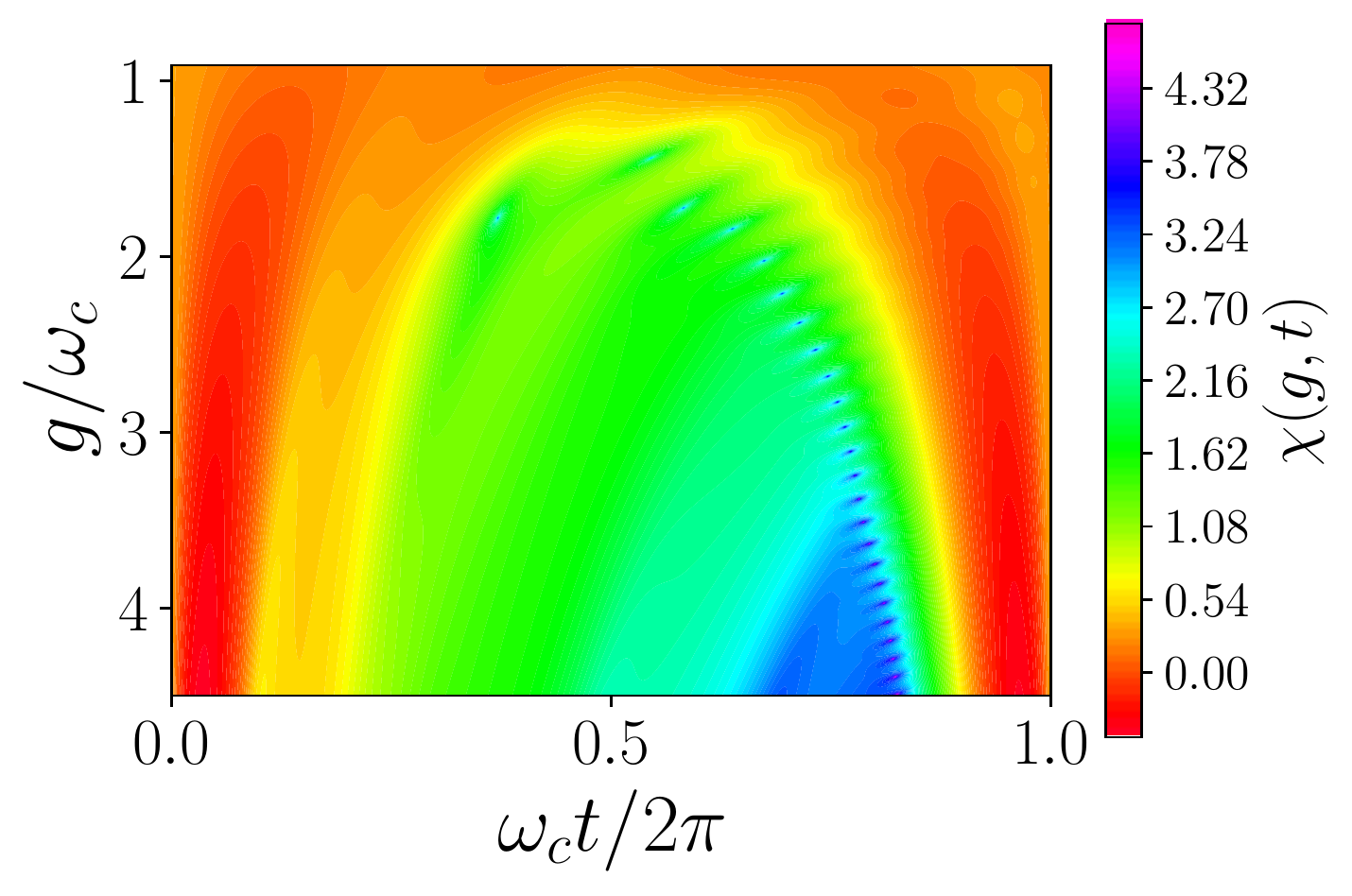}
	\caption{Values of $\chi(g,t)$ for the analyzed values of coupling strength and evolution time.}
	\label{fig:chi}
\end{figure}

Figs. \ref{fig:beta} and \ref{fig:chi} show plots of the behavior of $\beta(g,t)$ and $\chi(g,t)$, respectively, for the analyzed values of coupling strength and evolution time.

The radius of curvature for a parametric curve in two dimensions $\mathbf{r}(t)$ is given by
\begin{eqnarray}
	\kappa(t) = -\frac{[\mathbf{r}'(t) \times \mathbf{r}''(t)]_z}{\lVert \mathbf{r}'(t) \rVert^{3/2}},
	\label{eq:curvature}
\end{eqnarray}
where $[\mathbf{A}]_z$ is the $z$ component of $\mathbf{A}$. This is applied to our case by taking -- for a given value of the coupling strength -- the curve described by the following position vector:
\begin{eqnarray}
	\mathbf{r}(t) =
	\begin{pmatrix}
		\textrm{Re}(\mathcal{L}(g, t)) \\
		\textrm{Im}(\mathcal{L}(g, t))
	\end{pmatrix}.
\end{eqnarray}
It follows that the time derivatives of $\mathbf{r}(t)$ are given by the real and imaginary components of the time derivatives of the LA. If $\mathcal{L}(g,t)$ is expressed as
\begin{align*}
	\mathcal{L}(g,t) = A(g,t) \exp(i \theta(g,t)),
\end{align*}
then we have that
\begin{align*}
	\partial_t \theta(g,t) = \frac{[\mathbf{r}(t) \times \mathbf{r}'(t)]_z}{\lVert \mathbf{r}(t) \rVert^2}.
\end{align*}
Using a similar argument applied to $\partial_t \mathcal{L}$, we obtain that
\begin{align*}
	\kappa(t) = -\lVert \mathbf{r}(t) \rVert^{1/2} \partial_t \arg(\partial_t \mathcal{L}(g,t)).
\end{align*}
This means that the calculated chiralities will have the opposite sign to the defined curvature, giving a geometric interpretation of the results in Fig. \ref{fig:rotation}.

\section{Zero qubit splitting case} \label{app: ZQS}
In this $\omega_0\rightarrow 0$ limit, we consider the simplest quenching case in which the qubit-cavity coupling strength jumps suddenly from $0$ to a constant $g > 0$ at time $0$, and then maintains that constant value for time $t$. The operator
\begin{equation}\label{e5}
\hat{U}(t,0)=e^{i \omega_ct\hat{a}^{\dag}\hat{a}}e^{-\left [ \int_{0}^t ds K(s)\frac{dK^*(s)}{ds}\right ]}e^{i K(t)\hat{a}^{\dag}}e^{i K^*(t)\hat{a}},
\end{equation}
where
\begin{equation}\label{e6}
K(t)=\int_{0}^t ds \, g(s) e^{i \omega_cs}
\end{equation}
encodes the time variation of the qubit-photon coupling strength. Thus
\begin{eqnarray}\label{e12}
\hat{U}(t,0)|\psi(0)\rangle=e^{i \omega_ct\hat{a}^{\dag}\hat{a}}e^{i\theta(t)}\hat{D}\left ( iK(t) \right )|\psi(0)\rangle,
\end{eqnarray}
where $\hat{D}\left (\alpha \right )=e^{-\frac{|\alpha|^2}{2}}e^{\alpha \hat{a}^{\dag}}e^{-\alpha^* \hat{a}}$ is the displacement operator and $\theta(t)=\textrm{Im}\{ \int_{0}^t ds \, K(s)\frac{dK^*(s)}{ds} \}$. Hence
\begin{eqnarray}\label{e8}
\nonumber {\cal L}_0(t)&=&\langle \psi(0)|\hat{U}(t,0)|\psi(0)\rangle\\
&=&e^{i\theta(t)}\langle \psi(0)|e^{i \omega_ct\hat{a}^{\dag}\hat{a}}\hat{D}\left ( iK(t) \right )|\psi(0)\rangle.
\end{eqnarray}

Given an initial Fock state with $n$ photons, $|\psi(0)\rangle=|n\rangle$, this yields
\begin{eqnarray}\label{e88}
{\cal L}_0(t)=e^{i\theta(t)}e^{i n\omega_ct}e^{-\frac{1}{2}|K(t)|^2}L_n\left ( |K(t)|^2 \right ),
\end{eqnarray}
where $L_n\left ( x \right )$ is the Laguerre polynomial of order $n$.
Starting from the initial ground state $|\psi(0)\rangle=|g,0\rangle$ yields
\begin{eqnarray}\label{e89}
{\cal L}_0(t)=e^{i\theta(t)}e^{-\frac{1}{2}|K(t)|^2}.
\end{eqnarray}
which shows that the LA is the projection of the coherent state $|iK(t)\rangle=\hat{D}\left ( iK(t) \right )|0\rangle$ onto the initial vacuum state.
The Loschmidt decay factor $K(\tau)$ becomes
\begin{eqnarray}\label{e21}
K(t)=\frac{2g}{\omega_c}{\rm sin}\left ( \frac{\omega_c t}{2} \right )e^{i\frac{\omega_c t}{2}}
\end{eqnarray}
and so
\begin{eqnarray}\label{e22}
{\cal L}_0(g,t)=e^{i\theta(g,t)}e^{-2\left (\frac{g}{\omega_c}\right )^2{\rm sin}^2\left ( \frac{\omega_c t}{2} \right )},
\end{eqnarray}
with the phase
\begin{eqnarray}\label{e77}
\theta(g,t)=\left (\frac{g}{\omega_c}\right )^2\left [\omega_c t- {\rm sin}\left ( \omega_c t \right )\right ].
\end{eqnarray}

\section{Equation of motion for the LA} \label{app: EOM}
Using $\mathcal{L}(g,t) = \braket{\psi_0 | e^{-i \hat{H} t}| \psi_0}$, we obtain
\begin{eqnarray}
\frac{\partial \mathcal{L}}{\partial t} = -i \braket{\psi_0 | \hat{H} e^{-i \hat{H} t} | \psi_0}.
\end{eqnarray}
For the derivative with respect to $g$, an important aspect is that $\hat{H}$ and $d \hat{H}/dg$ do not commute which makes the computation of this derivative more complicated. Following previous results on the derivatives for exponential operators~\cite{Wilcox1967}:
\begin{eqnarray}
\frac{d e^{A(t)}}{d t} = \int_0^1 dx \, e^{xA(t)} \left( \frac{dA(t)}{dt} \right) e^{(1-x) A(t)}.
\end{eqnarray}
Applying this result to our case, we obtain
\begin{align}
&\frac{\partial \mathcal{L}}{\partial g} = 
&-i t \int_0^t dt' \, \braket{\psi_0 | e^{-i \hat{H} t'} \left( \frac{d \hat{H}}{d g} \right) e^{-i \hat{H} (t-t')} | \psi_0}.
\end{align}
Using the expression for the Hamiltonian given in Eq. \eqref{e1}, we obtain the following result for $g \ne 0$:
\begin{eqnarray}
-it \frac{d\hat{H}}{dg} = - i \frac{t}{g} \left( \hat{H} - \hat{N} + \frac{1}{2} \right).
\end{eqnarray}
Combining this with the result for $\partial \mathcal{L}/ \partial t$, we obtain
\begin{align}
\frac{\partial \mathcal{L}}{\partial g} =& \frac{t}{g} \frac{\partial \mathcal{L}}{\partial t} \\ \nonumber
&+ \frac{i}{g} \int_0^t dt' \, \braket{\psi_0 | e^{-i \hat{H} t'} \left( \hat{N} - \frac{1}{2} \right) e^{-i \hat{H} (t-t')} | \psi_0}.
\end{align}
Introducing a complete set of states after $\hat{N} + 1/2$ yields
\begin{align}
&g \frac{\partial \mathcal{L}}{\partial g} - t \frac{\partial \mathcal{L}}{\partial t} = \\ \nonumber
&+ i \sum_{n=0}^\infty \left[ \frac{(-1)^{n+1}}{2} + n \right] \int_0^t dt' \, \braket{\psi_0 | e^{-i \hat{H} t'} | n} \braket{n | e^{-i \hat{H} (t-t')} | \psi_0}.
\end{align}
Extracting the first term:
\begin{align}
&g \frac{\partial \mathcal{L}}{\partial g} - t \frac{\partial \mathcal{L}}{\partial t} = -\frac{i}{2} \int_0^t dt' \, \mathcal{L}(g,t') \mathcal{L}(g,t-t') \\ \nonumber
&+ i \sum_{n=1}^\infty \left[ \frac{(-1)^{n+1}}{2} + n \right] \int_0^t dt' \, \braket{\psi_0 | e^{-i \hat{H} t'} | n} \braket{n | e^{-i \hat{H} (t-t')} | \psi_0}.
\end{align}
Using Eq. \eqref{z66}, we obtain the following result:
\begin{align}
\mathcal{L}(g, t') \mathcal{L}(g, t-t') = & \mathcal{L}(g, t) 
- \sum_{n=1}^\infty \braket{\psi_0| e^{-i \hat{H} (t - t')} |n} \braket{n| e^{-i \hat{H} t'} |\psi_0}.
\end{align}
and hence the following, which leads to Eq. \eqref{Eq:z18}:
\begin{align}
&g \frac{\partial \mathcal{L}}{\partial g} - t \frac{\partial \mathcal{L}}{\partial t} = -\frac{i t}{2}\mathcal{L}(g,t) \\ \nonumber
&+ i \sum_{n=1}^\infty \left[ \frac{(-1)^{n+1} - 1}{2} + n \right] \int_0^t dt' \, \braket{\psi_0 | e^{-i \hat{H} t'} | n} \braket{n | e^{-i \hat{H} (t-t')} | \psi_0}.
\end{align}

\end{document}